\documentclass[oldversion]{aa} 
\usepackage{graphicx}
 \usepackage{lscape}
 \usepackage{colortbl}
 \usepackage{natbib}
 \usepackage{verbatim} 
  \usepackage{comment} 
\usepackage{ulem}
\usepackage{url}
 \bibpunct{(}{)}{;}{a}{}{,}
\usepackage{txfonts}
\begin{document}
\input psfig.tex

\titlerunning{Red sequence at $z\sim2$}

   \title{Red sequence determination of the redshift of the
   cluster of galaxies JKCS\,041:  $\mathbf{z\sim2.2}$} 
   \author{S. Andreon 
          \inst{1}
          \and
          M. Huertas-Company
	  \inst{2,3}
          }

   \institute{
             Observatorio Astronomico di Brera, via Brera 28, 20121, Milan, Italy\\
             \email{stefano.andreon@brera.inaf.it} 
	     \and
   GEPI, Paris-Meudon Observatory 5, Place Jules Janssen, 92190, Meudon, France\\
              \email{marc.huertas@obspm.fr}
         \and
             Universit\'e Paris Diderot, 75205 Paris Cedex 13, France\\
             }

   \date{Received --, 2010; accepted --, 2010}

\abstract{
This paper aims at robustly determining the redshift of the cluster
of galaxies JKCS\,041 and at putting constraints on the formation
epoch of the color-magnitude sequence in
two very high redshift clusters.
New deep $z'-J$ data show a clear narrow red
sequence that is co-centered with, and similary concentrated on,  
the extended X--ray emission of the cluster of galaxies JKCS\,041. 
The JKCS\,041 red sequence is  
$0.32\pm0.06$ mag redder in $z'-J$ than the red sequence of the
$z_{spec}=1.62$ IRC0218A cluster, putting JKCS\,041 at $z\gg1.62$
and ruling out $z \le 1.49$ the latter claimed by a recent paper.
The color difference of the two red sequences gives a red-sequence-based
redshift of $z=2.20\pm0.11$ for JKCS\,041, where the uncertainty accounts 
for uncertainties  in stellar synthesis population models, in photometric
calibration, and in the red sequence color of
both JKCS\,041 and IRC0218A clusters. 
We do not observe any sign of truncation of the red sequence 
for both clusters down to $J=23$ mag ($1 \ 10^{11}$ solar masses),
which suggests that it is already in place  in
clusters rich and massive enough to heat and retain hot gas 
at these high redshifts.
}

   \keywords{Galaxies: evolution --- galaxies: clusters: general 
   --- galaxies: clusters: individual JKCS\,041 
   --- galaxies: clusters: individual IRC0218A 
}

   \maketitle

\section{Introduction}
Many efforts have been made in past years to constrain the epoch
of formation of massive elliptical galaxies. Recent observational results
suggest that massive red-sequence galaxies were already assembled in cluster
cores at  $z\sim1.2$  (e.g. De Propris et al.
1999; Andreon 2006a; Lidman 2008) and possibly up to $z\sim2$
(Andreon 2010). Their colors and morphologies,
as revealed from observational studies in mass selected samples, remain
unchanged from $z\sim1$ (e.g. Holden et al. 2007; Huertas-Company et al. 2009), which
again suggests that these galaxies were assembled at even higher redshifts. 

These observations may contradict the predictions of
the $\Lambda$CDM model, in which mergers of gas-rich galaxies are the main
driver of spheroid formation (e.g. Toomre \& Toomre 1972) and the main
explanation for the over density of red galaxies in cluster cores (e.g.
Dressler 1984). Finding the cluster of galaxies at the highest
redshifts allows the epoch of formation of elliptical galaxies
to be pushed back, making it  a very intense activity in past years (e.g.
Papovich et al. 2010).

Recently, Andreon et al. (2009) detected what seems to be the most distant 
cluster of galaxies to date (JKCS\,041 at $z_{phot}=1.9$)
by looking at over densities in red galaxies. JKCS\,041
lies in the SWIRE/CFHTLS field at RA=2h, Dec=-4 $deg$ and benefits from an
extensive multi-wavelength follow-up, spanning 1.4 Ghz to X rays and
including SWIRE/Spitzer imaging in 7 IR bands (3.6, 4.5, 5.8, 8.0, 24, 70,
160 microns), as well as optical CFHTLS ugriz, WIRDS JHK, and UKIDSS JK 
data. 
The cluster has not been spectroscopically confirmed yet. However, in the
Chandra follow-up, it is detected as an extended X-ray source with $T\sim7.4$
keV, which confirms there is 
a hot intracluster medium, present in
formed clusters and lacking in protoclusters. 

In this paper we first report the presence of a clear red sequence of
passive galaxies in the region of JKCS\,041 using a filter pair 
sampling the Balmer break at $z>1.2$: the $z'-J$ color. 
Second, by comparing the red sequence color of the JKCS\,041 
and IRC0218A, a spectroscopically confirmed cluster at $z\sim1.62$ 
(Papovich et al. 2010: Tanaka et al. 2010), we confirm that JKCS\,041 lies at 
a much higher redshift,
and we quantify its photometric redshift ($z\sim 2.2$). 

Throughout the
paper, we assume the following cosmological parameters:
$H_0=70$ km s$^{-1}$ Mpc$^{-1}$, $\Omega_m=0.3$, and
$\Omega_\Lambda=0.7$. Magnitudes are in the AB system.

\begin{figure}  
\centerline{\psfig{figure=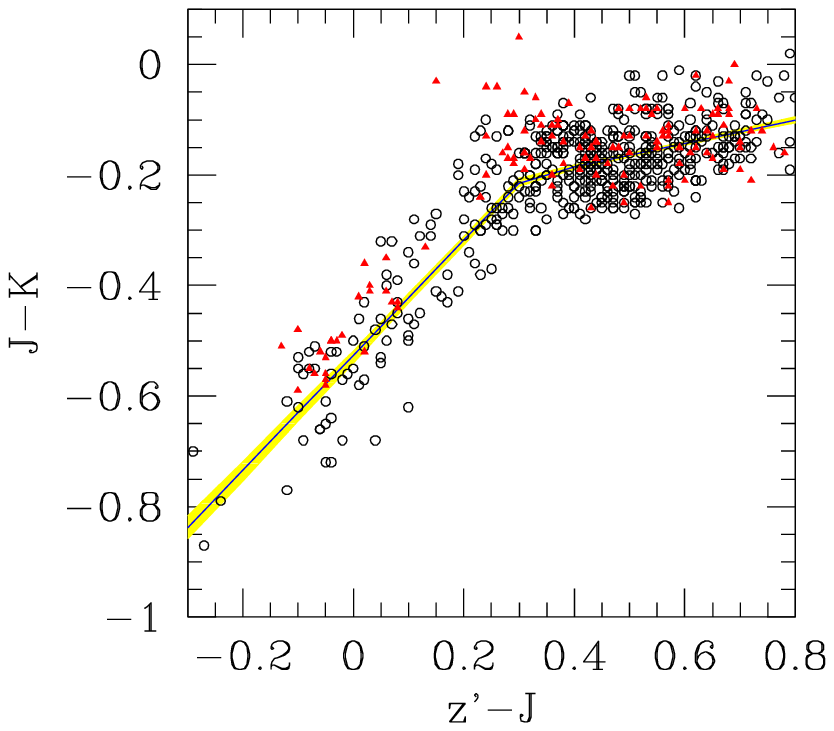,width=5truecm,clip=}}
\centerline{\psfig{figure=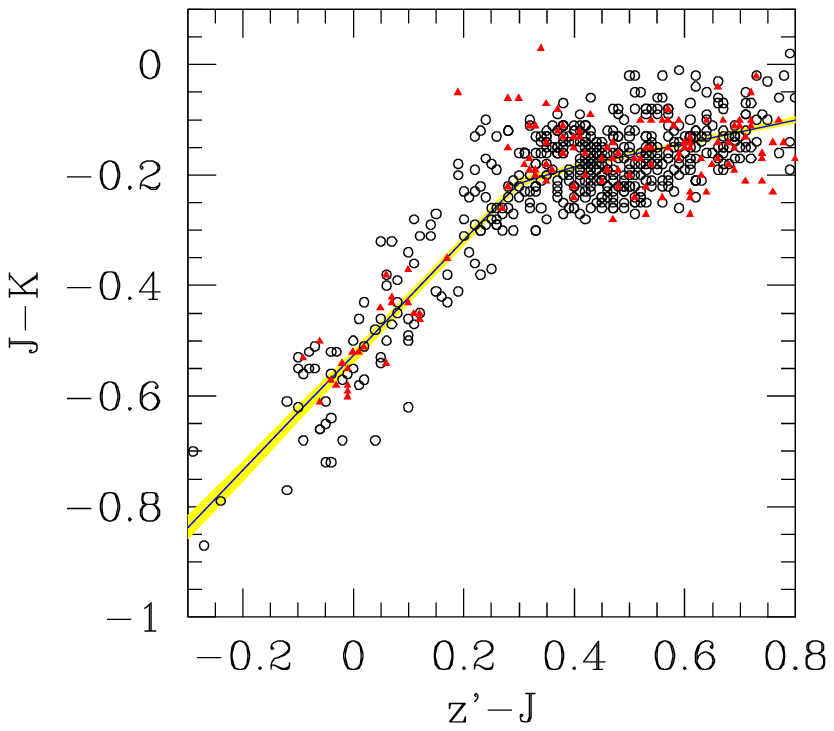,width=5truecm,clip=}}
\caption{Stellar locus from photometry as published (top panel) and 
after minor corrections to match locii determined
from different catalogs (bottom panel).  Red triangles and
open circles indicate stars in the direction of JKCS\,041 and IRC0218A, 
respectively.
The solid line shows the mean fitted locus. The shading
marks the 68 \% error (highest posterior interval) of the model.} 
\label{fig:stellarlocus} 
\end{figure}

\section{Data and minor photometric corrections} \label{sec:dataset} 

JKCS\,041 is in the area covered by  
CFHTLS deep survey 
and by  WIRDS follow-up in the infrared filters
($J, K$) (Bielby et al.,
in preparation, catalogs are available on the Terapix site). 
More precisely, we used the catalog generated using
$K$-band as detection image and the other bands ($z'$ and $J$)
in analysis mode.
We measured an $S/N\sim15$ at $z'=25$ ($2^{"}$ apertures used to measure 
colors) and $J=23$ mag.
The seeing in both filters is
very similar, i.e. $0.71^{"}$ FWHM in the $z'$ band (CFHTLS deep) and
$0.68^{"}$ in the J band (WIRDS), so no correction for seeing differences 
was applied. 

IRC0218A is in the area covered by  
Williams et al. (2009), whose catalogs are based on the
UKIDSS survey (Lawrence et al. 2007) for
the infrared bands ($J$ and $K$), and from the
SXDS survey (Furusawa 2008) 
for the $z'$ band. Colors listed in the catalog are 
measured in matched apertures and corrected for seeing
differences.
We measured an $S/N\sim15$ at $z'=25.2$ 
(in the apertures used to measure colors)
and $J=23$ mag.

\begin{figure}
\centerline{\psfig{figure=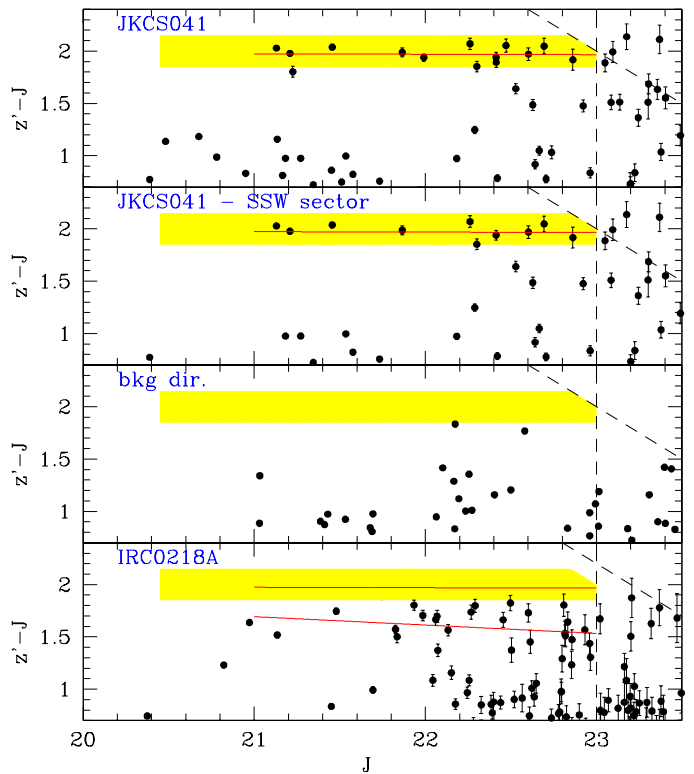,width=9truecm,clip=}}
\caption{Color--magnitude plots. Top panel: in the direction of JKCS\,041;
top-middle
panel: toward JKCS\,041 once the SSW octant has been flagged because
contaminated by a nearby structure, see text. Middle-bottom panel: 
in a control area of the same solid angle as the top panel;
bottom panel:  toward IRC0218A. 
Solid lines shows the fitted color-magnitude relations of
the two clusters, whereas shading indicates the simpler color range
$1.85<z'-J<2.15$. Slanted and vertical
dashed lines indicate magnitudes where the S/N=15 in $J$ (vertical)
or $z'$ (diagonal), and therefore delimit the region where catalogs should be
complete. JKCS\,041 red sequence 
galaxies are 0.32 mag redder than the red sequence of IRC0218A, at $z=1.62$,
indicating that JKCS\,041 is at $z\gg1.62$.
} 
\label{fig:CMRs}
\end{figure}

The photometry of the two catalogs comes from different telescopes 
and reduction pipelines.
An independent check that they are truly on a consistent
photometric system is therefore necessary.
The IRC0218A cluster lies 2.1 degrees away from JKCS\,041, and
both clusters are at high galactic latitude ($\sim 60$ deg). 
Stars in the two catalogs should, therefore, fall on the
same $z'-J$ vs $J-K$ locus. 
The top panel of Fig. 1 shows the stellar color-color locus in the
two directions: the two sequences almost coincide,
with a hard-to-notice small offset.
The offset (and its
error) is computed by a joint fit of the two stellar sequences
with a broken line, assuming uniform priors for the two line
position angles, for the $z'-J$ and $J-K$ of the break point, 
for the vertical thickness of the locus, and for the two color offsets.
The vertical thickness has a prior zeroed to negative
(meaningless) values. To account for outliers (most of which
are not visible in Fig 1 because outside the plotted color range, 
mainly QSO, variable stars and a few
objects with corrupted photometry), we model the vertical color spread
with a Cauchy distribution. The
larger wings of the Cauchy distribution de-weight outliers.
We found that stars in
the direction of JKCS\,041 are $0.039 \pm 0.014$ mag bluer
in $z'-J$, and $0.021\pm 0.007$ redder in $J-K$,
than in the direction of IRC0218A. We therefore corrected colors
listed in the WIRDS/CFHT catalog by
the computed photometric correction to make equal the 
stellar locii. The bottom panel
of Fig. 2 shows the stellar locii after 
the very small photometric correction.

\begin{figure}
\centerline{\psfig{figure=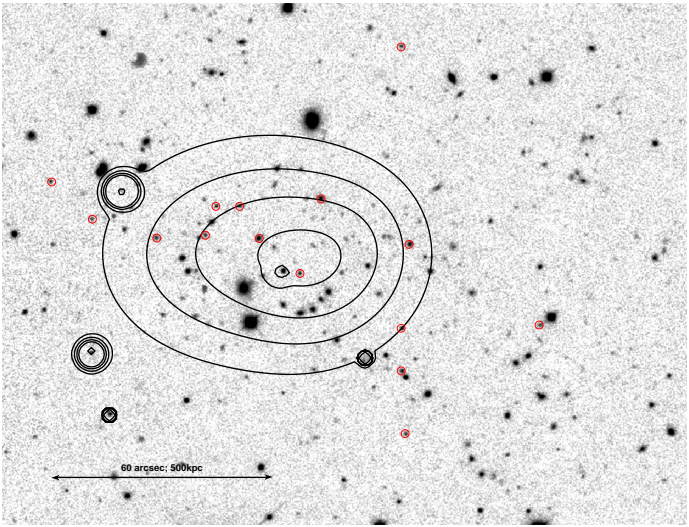,width=9truecm,clip=}}
\caption{$J$ band image of the field near JKCS\,041.
Contours indicate the adaptively smoothed X--ray emission detected by Chandra 
(from Andreon et al. 2009). Circles highligh galaxies on the red sequence
($1.85<z'-J<2.15$). These are concentrated and co-centered with 
the X--ray emission. North is up and east to the left. The ruler
is 1 arcmin wide ($\sim 500$ kpc at $z=2.2$).
} 
\label{fig:boh}
\end{figure}

\section{Results} 

\subsection{IRC0218A and JKCS\,041 red sequences} 

The top panel of  Figure~2
shows the color-magnitude relations toward JKCS\,041 within a
radius of $1$ arcmin ($\sim 0.5$ Mpc at $z\sim2$), 
after applying the minor photometric correction
described above. The SSW octant (i.e. the sector going from S to W
for 45 deg) is contaminated by galaxies of a nearby structure $\sim 1.2$
arcmin SSW discussed in sec 3.3. 
This octant is flagged in the top-middle panel. 
The middle-bottom panel 
shows 
a random control region of 1 arcmin radius for comparison. It turns
out to be at roughly 0.4 deg north of JKCS\,041. 
JKCS\,041 presents a clear red sequence of $\sim12$
well aligned objects, within a narrow $\sim 0.1$ color band, all within
1 arcmin from the X-ray cluster center. 

Inspection of the middle-bottom
panel shows no galaxy in the control field direction as red as in the 
JKCS\,041 direction. There is none in the shaded region, making 
the detection of JKCS\,041 very significant, a point already
shown (and quantified) in Andreon et al. (2009) using a control region 
with a much larger solid
angle, and it is readily visible here thanks
to the deeper WIRDS data.

Fig. 3 shows the spatial
distribution of galaxies within the shaded color range, $1.85<z'-J<2.15$, with 
no spatial filtering applied: these galaxies are not uniformly distributed
but concentrated in the region of the X-ray emission.

The bottom panel of Fig. 2 shows the color-magnitude relations toward
IRC0218A within a radius of $2$ arcmin ($\sim 1$
Mpc at $z=1.62$).  
The IRC0218A reddest galaxies are all bluer than the bluest galaxies on the
JKCS\,041 red sequence, and in fact none fall in the shaded area where
JKCS\,041 red sequence galaxies are found. 
The redder color of the JKCS\,041 red sequence directly implies
that $z\gg 1.62$. Even without any photometric correction,
the JKCS\,041 red sequence is 0.29 mag redder than the IRC0218A red sequence,
showing the much higher redshift of JKCS\,041 compared to $z=1.62$.

\begin{figure}
\centerline{\psfig{figure=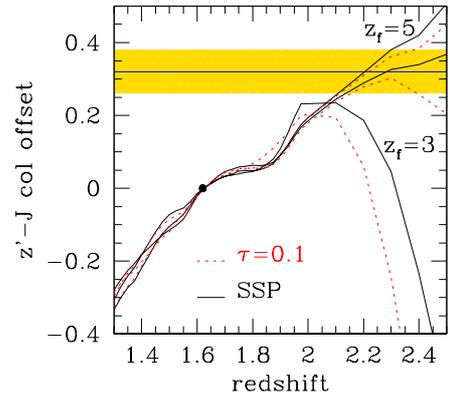,width=6truecm,clip=}}
\caption{Photometric estimate of the JKCS\,041 redshift. Color 
track of SSP (solid black) and $\tau=0.1$ (red dotted)
models with different formation
redshifts, $z_f=5,4,3$ (from top to bottom), are plotted, after
zero-pointing them to the observed color of IRC0218A cluster
at $z=1.62$. The measured color difference is indicated with an horizontal
line, and its error shaded. By simple eye inspection of this figure,
the JKCS\,041 photometric redshift is $z\sim 2.2$. 
} 
\label{fig:photoz}
\end{figure}

\subsection{JKCS\,041 photometric redshift} 

We fit the red sequence of JKCS\,041 and IRC0218A through the Bayesian
methods of Andreon (2006) and Andreon et al. (2006), also used for
other clusters (Andreon 2008; Andreon et al. 2008), solving
at once for all parameters (slope, intercept and intrinsic spread of the 
color-magnitude relation, and luminosity function parameters: characteristic
magnitude, faint end slope $\alpha$ and normalization). It also
accounts for photometric errors and the presence of
a fore/background population, the latter constrained using
a control field region with a $64$ times larger solid angle. 
The mean fitted red sequence is shown in
figure 2 at $z'-J\sim 2$ mag for JKCS\,041 and at $z'-J\sim 1.6$ mag 
for IRC0218A. We found the IRC0218A red sequence 
$0.32\pm0.06$ mag bluer than the JKCS\,041 red sequence. The quoted
error accounts for the error in the two red sequence intercepts 
($0.035$ for JKCS\,041 and $0.05$ mag for IRC0218A) and
the uncertainty in the photometric correction ($0.014$ mag, see sec 2).

Fig. 4 plots color 
tracks of single stellar populations (SSP) and  
exponential declining ($\tau=0.1$) models with solar
metallicity, Chabrier initial mass function, formed at $z_f=5,4,3$, 
zero-pointing them to the observed color of IRC0218A cluster
at $z=1.62$.  This approach to estimating the
photometric redshift is more robust 
than using the absolute color of
the population and benefits from a smaller extrapolation. 
For computing the color tracks we used the 2007 version of
Bruzual \& Charlot (2003) synthesis population model.
The JKCS\,041 photometric redshift derived
from an SSP model with
$z_f=5$ can be read easily in it:
$z_{phot}=2.20 \pm 0.10$. We add an additional 0.05 error term
in quadrature 
to approximatively account for different $z_f$ and star formation histories
(accounting for the $0.07$ total spread in redshift of the
three models reaching a color difference of $0.32$ mag),
giving a final photometric redshift of $z_{phot}=2.20 \pm 0.11$.
Finally, we note that if models with $z_f=4$ and $5$
and an unlikely high $2.5\times$ solar metallicity 
were adopted, then the estimated redshift would
be only marginally different, about 0.05 lower.

The newly determined photometric redshift is slightly higher than the
conservatively estimated photometric redshift quoted
in Andreon et al. (2009)
and consistent with it, since the two 68 \% confidence intervals 
([1.84,2.12] vs [2.09,2.31]) overlap. 
The current redshift determination supersedes
the old determination since we now use a calibration at similar
redshift (via IRC0218 red sequence color).

Fig. 4 also puts constraints on the formation redshift of
the stellar population: $z_f=3$ is excluded at 90 (97) \% confidence level
for an SSP ($\tau=0.1$) model of solar metallicity.

The estimated redshift of JKCS\,041 puts the cluster at an epoch when 
major changes to the color--magnitude relation
are expected to occur, according to theoretical predictions. 
While theoretical studies (e.g.
Menci et al. 2008) claim that clusters at redshift about $2.0$ 
are in a phase where the bright end of the red sequence has yet not been
formed, the top-middle panel of Fig. 2 shows that
the JKCS\,041 red sequence is already built, indicating that (i)
star formation in massive galaxies should be halted well before, or
more efficiently, than in current theoretical modeling of galaxy evolution
and (ii) 
that we still have not reached the epoch when the brightest cluster galaxies were
actively forming stars, and hence were blue.

Finally, we briefly discuss the other end of the red
sequence. In both JKCS\,041 and IRC0218A
we do not observe any sign of truncation of the red sequence down to 
$J=23$ mag ($1\ 10^{11}$ solar masses for an SSP with $z_f=5$),
which suggests that it is already in place at these
high redshifts in clusters that are rich and massive
enough to heat and retain an hot gas. The slightly different claim
for IRC0218A by Papovich et al. (2010) and Tanaka et al. (2010) concerns
a deficit at magnitudes
fainter than our $S/N=15$ limit, where we deem these catalogs
incomplete.
Deeper observations and a quantitative analysis are needed to
settle the issue at masses lower than $1\ 10^{11}$ 
solar masses.

\subsection{A lower redshift/different identification for JKCS\,041?}

\begin{figure}
\centerline{\psfig{figure=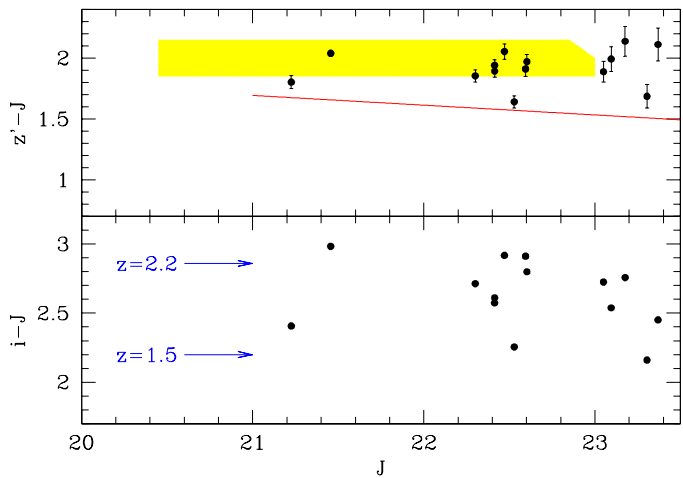,width=9truecm,clip=}}
\caption{Color--magnitude plot in the direction of JKCS\,041. Points
show red sequence galaxies with $z_{phot}=1.49$ according to Bielby et al.
(2010). {\it Top panel:} The solid line indicates the red sequence of the $z_{spect}=1.62$
IRC0218A cluster, and it is 0.35 mag bluer, on average, than points
{\it Bottom
panel:} Predicted (arrows) $i-J$ red sequence color at $z=1.5$ and $z=2.2$  
and observed (points) $i-J$ color of JKCS\,041 galaxies on the red sequence. 
} 
\label{fig:bielby}
\end{figure}

Bielby et al. (2010) 
propose
a different photometric redshift, $z=1.49$
for much the same galaxies we have plotted in the top panel of Fig 2, using
the 
same catalog as we do, the  excellent one produced by Bielby et al. (in
preparation). 
More precisely, Fig. 5 plots the color-magnitude diagram of the galaxies
these authors claim to be at $z_{phot}\sim1.49$ and on the red sequence. 
We got coordinates from the authors
and we plot (their and our) photometry in Fig 5. These galaxies are redder
than the IRC0218A $z=1.62$ red sequence, by $0.35$ mag on average
(median), whereas 
if they were red sequence $z\sim1.49$ redshift galaxies, as Bielby et al. (2010) 
claim, they should be 0.15 mag bluer (see Fig. 4),
not about 0.4 mag redder. 

The bottom panel of Fig. 5 shows that a similar
conclusion also holds for the $i-J$ color displayed by Bielby et al. (2010)
in their Fig 10: JKCS\,041 red sequence galaxies have an $i-J$ color 
appropriate
at $z=2.2$, and are 0.34 mag too red at $z=1.49$.
The correctness of the template $i-J$ color is confirmed by the
observed color, $i_{775}-J\sim2.3$ mag, of the red sequence of 
the cluster XMMU J2235-2557
at $z=1.39$ (Strazzullo et al., 2010).

As briefly mentioned, the SSW octant of JKCS\,041 is contaminated by
bright and blue galaxies ($J\sim21$ and
$z'-J\sim1$),  as
an attentive inspection of Fig. 2 shows.
Fig. 6 shows the spatial
distribution of these galaxies with no spatial filtering
applied: they are concentrated at slightly more than $1$ arcmin 
southeast of 
JKCS\,041 and there is no reason for them
to be the optical counterpart of the extended
X--ray emission, owing to the too large angular offset, and because another
galaxy overdensity is spatially aligned with the X-ray emission. The
presence of another cluster at 1' is unsurprising,
given the observed average density
of about 210 (optical) clusters and groups per square degree
measured in the COSMOS region by Finoguenov et al. (2007):
there is a $\sim 20$
per cent probability of finding one of them, just by chance, within 1' 
of any sky position.
The contamination of this offset group in the solid angle 
used by Bielby et al. (2010)
to measure the distribution of photometric redshifts toward
JKCS\,041 is the likely
reason for their  
suggestion that JKCS\,041 may be at $z\sim1.1$.
Actually, the spatial
offset between the X-ray emission and the galaxies originating
the $z\sim1.1$ photometric redshift (much of the same galaxies as in our
Fig. 6) is already evident in their figure 10. 

\begin{figure}
\centerline{\psfig{figure=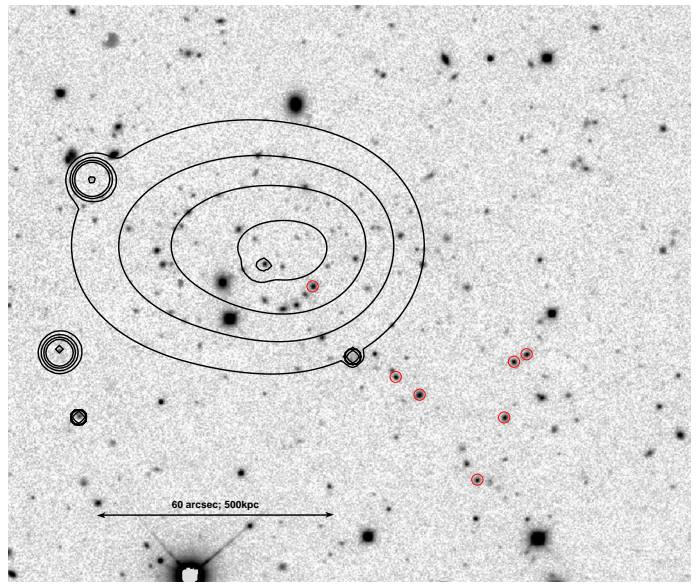,width=9truecm,clip=}}
\caption{$J$ band image of the field near to JKCS\,041.
Contours show the adaptively smoothed X-ray emission detected by Chandra 
(from Andreon et al. 2009). Circles mark blue and bright galaxies 
($J\sim 21$ and $z'-J\sim 1$). They are offset by 1 arcmin SE from the 
X--ray emission. North is up, and east to the left. The ruler
is 1 arcmin wide.
} 
\label{fig:boh2}
\end{figure}

\section{Conclusions}

We show that galaxies $0.32$ mag redder than the red sequence of
the $z_{spec}=1.62$ cluster IRC0218A are spatially concentrated
where the JKCS\,041 X-ray emission is located. This implies $z\gg 1.62$
and rules out $z \le 1.49$ the latter claimed by a recent paper. 
The 0.32 mag color difference of the 
two red sequences 
implies that the cluster JKCS\,041 is at $z=2.20\pm0.11$, 
where the uncertainty accounts 
for uncertainties  in stellar synthesis population models, in photometric
calibration, and in the red sequence color of
both JKCS\,041 and IRC0218A clusters.

We can thus confirm that JKCS\,041 is a cluster of galaxies with 
the photometric redshift $z_{red \ sequence}=2.20\pm0.11$, with a formed
potential well, deep enough to be hot and retain the intracluster medium,
and with a well-defined red sequence. 

Incoming X--ray survey telescopes or red-sequence based surveys
will likely return hundreds of $z\sim2$ cluster candidates.
Getting spectroscopic redshifts for all of them, or even a small part, 
is too time-consuming with current telescopes. Therefore,
photometric redshifts based on the red sequence color will necessarily become
very popular in the next years, so we need to get used to them.

\begin{acknowledgements} We thank Rich Bielby for sending us
coordinates for galaxies marked in their Fig. 10 and  
Henry McCracken for useful information on the WIRDS catalog.
We thank Rich Bielby, Ginevra Trinchieri, Roberto De Propris,
and Peter Edmonds for comments on an earlier version of this paper.
Based on observations obtained with
MegaPrime/MegaCam\footnote{The full text acknowledgement is at
http://www.cfht.hawaii.edu/Science/CFHLS/cfhtlspublitext.html} and 
WIRCAM\footnote{The full text acknowledgement is  at
http://ftp.cfht.hawaii.edu/Instruments/Imaging/WIRCam/WIRCamAcknowledgment.html}
at CFHT.

\end{acknowledgements}

{}

\end{document}